\documentclass[10pt]{article}
\usepackage[latin1]{inputenc}
\usepackage{amsmath}
\usepackage{amsfonts}
\usepackage{amssymb}
\usepackage{graphicx} 
\begin{document}
\begin{center}
{\bf DENSITY FLUCTUATIONS  IN THE OSCILLATORY PHASE OF NONCLASSICAL INFLATON  IN  FRW UNIVERSE
}
\\[.4 cm]
K.K.Venkataratnam\footnote{email:kkvratnam@yahoo.com} and P.K. Suresh\footnote{ pkssp@uohyd.ernet.in}\\[.4cm]

{ \it School of Physics,University of Hyderabad,\\
Hyderabad-500046.India.
}
\end{center}

\begin{abstract}
Using coherent and  squeezed state formalisms of quantum optics for a minimally coupled non-classical inflaton in the FRW mertic is studied, in semiclassical theory of gravity. The leading order solution for the semiclassical Einstein equations  in the coherent, squeezed  and squeezed vacuum states  are obtained perturbatively and are exhibit powerlaw expansion behaviour. The validity of the semiclassical theory is examined in the squeezed vacuum state in the oscillatory phase of the inflaton. The semiclassical theory  in the oscillatory phase of the non-classical inflaton holds only if the associated squeezing parameter is much less compared to unity. Quantum fluctuations of the inflaton is also examined in coherent and squeezed state formalisms.
\end{abstract}

\section{Introduction}	
In standard cosmology,  description of  early universe is  based on the
Friedmann equations with  scalar field(s). The
Friedmann equations are  based on classical gravity  and  assume that  it valid even at
very early stage of the universe also. However, quantum effects  and quantum fluctuations of matter fields
are expected to play a significant role in this regime, though quantum gravity
effects are still negligible. To describe very early universe both gravity and matter fields are to be treated quantum mechanically, but at present there is no consistent quantum theory to describe  gravity. Therefore,  proper description of  early  universe  with an appropriate cosmological model can be studied in terms of the semiclassical Friedmann equations, where  gravity can be treated as classical with quantised matter field(s).

Recently, it has been found that non-classical state formalisms of quantum optics are found quite useful to deal with quantum effects in
cosmology. The  inflationary scenario  has
received much attention  in semiclassical theory of gravity \cite{1} \cite{5}. There are works in which  quantum properties of
the inflaton, which is responsible for the inflation, were investigated by many authors\cite{6} \cite{15}.
Such  studies show that results obtained in classical gravity are
quite different from those in semiclasical gravity. Though both classical and quantum
inflaton in the oscillatory phase  lead  the same power law
expansion, the correction to  expansion does not show any oscillatory
behaviour in semiclassical gravity in contrast to the oscillatory behaviour
seen in classical gravity. It is to be  noted that, the coherently oscillating inflaton
suffers from particle production. Such studies reveal that quantum
effects  and quantum phenomena  play an important role in inflation scenario and  related issues.

Recently, we have studied\cite{16}  particle production during the oscillatory phase of the inflaton in squeezed and coherent state formalisms of quatum optics, in semiclassical theory of gravity. It is also interesting to study the validity of the semiclassical Einstein equation in the oscillatory phase of inflaton, because of quantum nature of the particle creation which may fluctuates the corresponding energy density so widely. It is noted that the semiclassical theory of the Einstein's  equations are valid only if the fluctuations of energy momentum tensor  are not too large. The present work is to study the validity of the semiclassical Einstein equation in squeezed vacuum state during the oscillatory phase of inflaton. The validity of the equation can be studied with the help of  the fluctuations of energy momentum tensor of the quantised inflaton.

The primary goal of the  paper is to study quantum fluctuations and density fluctuations of a  massive, minimal inflaton in the FRW universe in the  semiclassical gravity by representing the inflaton in
squeezed vacuum in the oscillatory phase
of the inflaton. Hence to examine the validity of the Einstein equation  in the oscillatory phase of the inflaton.
Also we examine whether the solutions of the semiclassical equation, in the oscillatory phase of inflaton in coherent state, squeezed vacuum state and squeezed state, can lead to powerlaw expansion behaviour.
\section{Semiclassical Einstein equation and inflaton in FRW metric}
Most of the cosmological models are based on the classical gravity of
Einstein equations with scalar field  on the FRW metric. To study
scalar field and the corresponding Friedmann equations at a deeper level, both background
metric and the field are to be treated  quantum mechanically. Since a consistent
quantum theory of gravity is not available, in most of the  cosmological models,
 the back ground metric is considered as classical (not quantised) and matter
field as quantum. Such an approximation of the Einstein equation is known as
semiclassical Einstein equation\cite{17}. Therefore, in  a semiclassical theory of
gravity, Einstein equations take the following form
(here onwards we use the unit system
 $\hbar=c=1$ and $G=\frac{1}{m^{2}_{p}}$):
 \begin{equation}
G_{\mu\nu}=\frac{8\pi}{m^{2}_{p}} \langle T_{\mu\nu}\rangle\,.
\end{equation}
Here, $G_{\mu\nu}$ is the Einstein tensor  and
$\langle T_{\mu\nu}\rangle$  is the expectation value of the energy-momentum
tensor for matter field in a suitable quantum state under consideration with the
quantum state satisfying the Schr${\ddot{o}}$dinger equation:
\begin{equation}
   i \frac{\partial}{\partial t}\mid \psi\rangle =\hat{H}\mid \psi\rangle\,\,.
 \end{equation}
 $\mid\psi\rangle$ denotes a quantum state and
$ \hat{H}$ is the Hamiltonian operator governing the quantum state.
   The semiclassical theory of gravity provides a description of the
 gravitational field of quantum systems with the expectation value of
energy-momentum tensor as the source of gravity.
 Consider a flat Friedmann-Robertson-Walker  spacetime
 \begin{equation}
 ds^{2}=-dt^{2}+S^{2}(t) (dx^{2}+dy^{2}+dz^{2}).
 \end{equation}
Assume that a massive inflaton (a homogeneous scalar field) field, minimally
coupled with gravity  and its Lagrangian density can be written as
\begin{equation}
{\cal{L}} = - \frac{1}{2} \sqrt{(-g)} ( g^{\mu \nu} \partial_\mu  \phi \partial_\nu \phi + m^2 \phi^2 ) .
\end{equation} 
The equation  of motion of the inflaton is governed by  the Klein-Gordon equation and
can be written by using eqns (3) and  (4) as
\begin{equation}
  \ddot\phi+3\frac{\dot{S}(t)}{S(t)}\dot\phi+m^{2}\phi = 0,
\end{equation}
where overdot represents a derivative with respect to time.
The inflaton field can be quantised by defining momentum conjugate to $\phi$ as
$\pi = \frac{\partial L}{\partial\dot{\phi}}$
and  following the canonical quantisation procedure, the Hamiltonian for inflaton can be written as
\begin{equation}
H=\frac{\pi^{2}}{2S^{3}(t)}+{1\over2}S^{3}(t) m^{2}\phi^{2}\,\,.
\end{equation}
Therefore, the temporal component of the energy-momentum tensor for the
inflaton is obtained as
\begin{equation}
T_{00}=S^{3}(t)({1\over2}\dot\phi^{2}+{1\over2} m^{2}\phi^{2})\,\,.
\end{equation}
\section{Coherent states and squeezed states}
Coherent and squeezed states\cite{18}\cite{20}are important classes of quantum states,
well-known in quantum optics. These states are being used as probes for
studying  the quantum effects in cosmology. Coherent states are considered  as
most classical, that can be generated from the vacuum state $\mid 0 \rangle$ by
the action of displacement operator. In the present study, we use single mode
coherent and squeezed states only. A single mode coherent state can be defined as
\begin{equation}
\mid\alpha\rangle=D(\alpha)\mid 0\rangle\,,
\end{equation}
where $D(\alpha)$ is the single mode displacement operator,  given by
\begin{equation}
D(\alpha)=\exp({\alpha a^{\dag}-\alpha^{*}a})\,\,.
\end{equation}
Here, $\alpha$ is a complex number and $a$, $a^{\dag}$ are respectively the
annihilation and creation operators, satisfying $[a,a^{\dag}]=1$. The action
of $a$  on the
coherent state gives
\begin{eqnarray}
a\mid\alpha\rangle=&\alpha\mid\alpha\rangle.
\end{eqnarray}
The single mode displacement operator given by (9) satisfy the following
property.
\begin{eqnarray}
D^{\dag} a D=&a+\alpha.
\end{eqnarray}
A squeezed state is generated by the action of the squeezing operator on any
coherent
state is also  on the vacuum state. Therefore, a single mode squeezed
state is defined as
\begin{equation}
\mid \alpha,\xi \rangle = Z(r,\varphi) D(\alpha)|0 \rangle\,\,,
\end{equation}
with $Z(r,\varphi)$, the
single mode squeezing operator   given by,
\begin{equation}
Z(r,\varphi)=\exp{r\over2}(e^{-i\varphi} a^{2}-e^{i\varphi} a^{\dag 2})\,\,.
\end{equation}
Here,  $r$ is the squeezing parameter, which
determines the strength of the squeezing and $\varphi$ is the squeezing
angle, which
determines the distribution between conjugate variables,
with  $0\leq r\leq \infty$ and $-\pi\leq\varphi\leq\pi$.
The  squeezing  operator satisfy the following property
\begin{eqnarray}
Z^{\dag}a Z =& a \cosh r-a^{\dag}e^{i\varphi}\sinh r  .
\end{eqnarray}
By setting $\alpha$=0 in  (12), one obtains the squeezed vacuum state, and is defined  as 
\begin{equation}
\mid\xi\rangle=Z(r,\varphi)\mid 0\rangle\,\,.
\end{equation}
The squeezed vacuum state is  a many-particle state and hence
the resulting field may be called classical. However, the statistical
properties of these states
greatly differ from the coherent states and therefore, it is
considered as highly non-classical  having no analog in classical physics.
In the case of coherent states, the variance of the conjugate variables are
always equal to each other, while in a squeezed state one component of the noise
is always squeezed with respect to the other. Therefore, in  (x,p) plane, the
noise for the coherent state can be described by a circle and for the
 squeezed state, it as an ellipse.
\section{Oscillatory phase of  inflaton and power law expansion}
In this section, we consider a massive inflaton minimally coupled to a
spatially
flat FRW metric and study  inflaton in  nonclassical states.
 First, we consider the inflaton
  in a nonclassical state which is closest to the  classical
 state i.e., the coherent states and
 then extend it to the  squeezed state representation of the inflaton.

 As mentioned earlier, we consider a massive inflaton, minimally coupled to a spatially flat
FRW universe with the metric (3). Therefore, the
time-time component of the classical gravity is now the classical Einstein equation
\begin{equation}
\left(\frac{\dot{S}}{S}\right)^2=\frac{8\pi}{3m^{2}_{p}} \frac{T_{00}}{S^3(t)},
\end{equation}
where $T_{00}$ is the energy density of the inflaton, given by (7).

In  the cosmological context, the classical Einstein equation (16) means that
the Hubble parameter, $H=\frac{\dot{S}}{S}$, is determined by the energy density of
the dynamically evolving inflaton as described by (16).
In the semiclassical theory, the Friedmann equation can be written as:
\begin{equation}
\left(\frac{\dot{S}}{S}\right)^2=\frac{8\pi}{3m^{2}_{p}} \frac{1}{S^3(t)}    
\langle \hat{H}_m \rangle,
\end{equation}
where $\langle \hat{H}_m \rangle$ represent the expectation value of the Hamiltonian
of the scalar field in a quantum state under consideration.

The massive minimal inflaton in the flat FRW universe can be described by the
time dependent harmonic oscillator  with the Hamiltonian given by eq(6). For
the semiclassical Friedmann equation (17), we have to find the expectation
value of the  Hamiltonian in a given quantum state. Then the eigenstates of
the  Hamiltonian are the Fock states:
\begin{equation}
\hat{a^{\dag}}(t)\hat{a}(t)|n,\phi,t \rangle = n|n,\phi,t \rangle,
\end{equation}
where $a$ and $a^{\dag}$ are the  annihilation and creation operators obeying
boson commutation relations $[a,a^{\dag}]=1$, the other combinations being
zero. These can respectively be written as:
\begin{eqnarray}
\hat{a}(t)&=&\phi^*(t) \hat{\pi} - S^{3}\dot{\phi}^*(t)
\hat{\phi},\nonumber\\
\hat{a^{\dag}}(t)&=&\phi(t) \hat{\pi}-S^{3} \dot{\phi}(t)
\hat{\phi}.
\end{eqnarray}
The expectation value of the Hamiltonian can  be calculated in the number
state by using (17), (18) and (19) and
hence, the semiclassical Friedmann equation (17), in the number state, can be written as
\begin{equation}
\left( \frac{\dot{S}(t)}{S(t)}\right)^{2}=\frac{8\pi}{3m^{2}_{p}}\left[
(n+{1\over2})(\dot{\phi}^*\dot{\phi }+m^{2}\phi^*\phi)\right].
\end{equation}
In  (20),  $\phi$ and $\phi^*$ satisfy (5) and the
Wronskian condition 
\begin{equation}
S^3(t)\left(\dot{\phi^*}(t)\phi(t)-\phi^{*}(t) \dot{\phi}(t) \right)=i.
\end{equation}
The Wronskian and the boundary conditions, fix the normalisation constants
of the  two independent solutions.

As an  alternative to the $n$  representation of the inflaton,  we next consider
the inflaton  in the coherent  and the squeezed states formalisms
and hence the semiclassical Einstein equation can be expressed in terms of
their respective parameters.

Using (8-11), (17) and (19)  the semiclassical  Einstein
equation can be written for the coherent state as
\begin{equation}
\left( \frac{\dot{S}(t)}{S(t)}\right)^{2}=\frac{8\pi}{3m^{2}_{p}}\left[
(|\alpha|^{2}+{1\over2})(\dot{\phi}^*\dot{\phi}+m^{2}\phi^*\phi)-{1\over2}
\alpha^{*2}(\dot{\phi}^{*2}+m^{2}\phi^{*2})
-{1\over2}\alpha^{2}(\dot{\phi}^{2}+m^{2}\phi^{2})\right].
\end{equation}
By using (12-14), (17)  and  (19),
  the semiclassical equation  for the  squeezed vacuum  state is obtained as
\begin{eqnarray}
\left(\frac{\dot{S}(t)}{S(t)}\right)^{2}&=&\frac{8\pi}{3m^{2}_{p}}\left[
(\sinh^{2}r+{1\over2})(\dot{\phi}^{*}\dot{\phi}+m^{2}\phi^{*}\phi)\right.
\nonumber\\
&&+\left.\frac{\sinh 2r}{4}\left[e^{-i\varphi}(\dot{\phi}^{*2}+m^{
2}\phi^{*2})
+e^{i\varphi} (\dot{\phi}^{2}+m^{2}\phi^{2})\right]\right].
\end{eqnarray}
Similarly, the semiclassical equation for a squeezed state is obtained as
\begin{eqnarray}
\left(\frac{\dot{S}(t)}{S(t)}\right)^{2}&=&\frac{8\pi}{3m^{2}_{p}}
\left[(\sinh^{2}r+{1\over2}+|\alpha|^{2})(\dot{\phi}^{*}\dot{\phi}
+m^{2}\phi^{*}\phi) \right.\nonumber\\
&&+\left(\frac{e^{-i\varphi}\sinh r\cosh r-\alpha^{*2}}{2}\right)
(\dot{\phi}^{*2}+m^{2}\phi^{*2}) \nonumber\\
&&\left.+\left(\frac{e^{i\varphi}\sinh r\cosh r-\alpha^{2}}{2}\right)
(\dot{\phi}^{2}+m^{2}\phi^{2})\right].
\end{eqnarray}
Our next task is to solve these semiclassical Einstein equations
for which  we transform the solution in the following form
\begin{equation}
\phi(t)=\frac{1}{S^{3\over2}}\psi(t),
\end{equation}
thereby obtaining
\begin{equation}
\ddot{\psi}(t)+\left(m^{2}-{3\over4}\left(\frac{\dot{S}(t)}{S(t)}\right)^{2}
-{3\over2}
\frac{\ddot{S}(t)}{S(t)}\right)\psi(t)=0.
\end{equation}
Now  we focus on the oscillatory phase of the inflaton after inflation. In the
parameter region satisfying  the inequality
\begin{equation}
m^2 > \frac{3\dot{S}^2}{4S^2}+\frac{3\ddot{S}}{2S},
\end{equation}
the inflaton has an oscillatory solution of the form
\begin{equation}
\psi(t)=\frac{1}{\sqrt{2w(t)}}\exp(-i\int w(t)dt),
\end{equation}
with
\begin{equation}
w^{2}(t)=m^{2}-{3\over4}\left(\frac{\dot{S}(t)}{S(t)}\right)^{2}-{3\over2}
\frac{\ddot{S}(t)}{S(t)}+{3\over4}\left(\frac{\dot{w}(t)}{w(t)}\right)^{2}-
{1\over2}\frac{\ddot{w}(t)}{w(t)}.
\end{equation}
  Using (25),(28) and (29)  in (20), we get
\begin{equation}
\left(\frac{\dot{S}(t)}{S(t)}\right)^2 = \left[\frac{4\pi}{3m^{2}_{p}}
(n+{1\over2})\frac{1}{S^3}\frac{1}
{w(t)}
\left(w^{2}(t)+m^{2}+{1\over4}
\left(\frac{\dot{w}(t)}{w(t)}+\frac{3\dot{S}(t)}{S(t)}\right)^2
\right)\right].
\end{equation}
Therefore
\begin{equation}
S(t) = \left[\frac{4\pi}{3m^{2}_{p}}(n+{1\over2})\frac{1}
{w(t)(\frac{\dot{S}(t)}{S(t)})^2}
\left(w^{2}(t)+m^{2}+{1\over4}
\left(\frac{\dot{w}(t)}{w(t)}+\frac{3\dot{S}(t)}{S(t)}\right)^2
\right)\right]^{1\over3}.
\end{equation}
 We solve  (31)  using the  following approximation
ansatzs
\begin{equation}
w_{0}(t)=m; \, \, \, \, S_{0}(t)=S_{0}t^{2\over3}.
\end{equation}
 Thus the next order perturbative solution is obtained for the number state as
\begin{equation}
S_1(t)=\left[\frac{6\pi}{m^{2}_{p}}(n+{1\over2})mt^2\left(1+\frac{1}{2m^2t^2}
\right)\right]^{1\over3}.
\end{equation}
Similarly the perturbative solution for the coherent state can be computed
  and  hence the scale factor for the  coherent state is obtained as
\begin{eqnarray}
S(t) &=&\left[\frac{8\pi}{3m^{2}_{p}(\frac{\dot{S}(t)}{S(t)})^{2}}
\left[
\frac{\left(|\alpha|^{2}+{1\over2}\right)(A+w^{2}(t)+m^{2})}{2w(t)}
\right.\right.\nonumber\\
&&-{1\over2}\frac{\alpha^{*2}\exp(2i\int w(t)dt)}{2w(t)}\left(A-w^{2}(t)-iw(t)
\left[3\frac{\dot{S}(t)}{S(t)}+\frac{\dot{w}(t)}{w(t)}\right]+m^{2}\right)\nonumber\\
&&-\left.\left.
{1\over2}\frac{\alpha^{2}\exp(-2i\int w(t)dt)}{2w(t)}
\left(A-w^{2}(t)+iw(t)
\left[3\frac{\dot{S}(t)}{S(t)}+\frac{\dot{w}(t)}{w(t)}\right]+m^{2}\right)
\right]\right]^{1\over3},
\nonumber
\end{eqnarray}
where
\begin{equation}
A={1\over 4} \left(\frac{\dot{w}}{w}+3\frac{\dot{S}}{S}\right)^2.
\end{equation}
Applying  the approximation ansatzs (32), then
the next order perturbative solutions for the coherent state is obtained as
\begin{eqnarray}
S_{1}(t)&=&\left[\frac{6 \pi t^{2}}{m^{2}_{p}}\left[\frac{(|\alpha|^{2}+
{1\over2}) (\frac{1}{t^{2}}+2m^{2})} {2m} \right.\right.\nonumber\\
&&-\frac{1}{2}\frac{\alpha^{*2}\exp(2i\int m dt)}{2m}\left(\frac{1}{t^{2}}-
\frac{2im}{t}\right)\nonumber\\
&&-\left.\left.\frac{1}{2}\frac{\alpha^{2}\exp(-2i\int m dt)}{2m} \left(\frac{1}{t^{2}}
+\frac{2im}{t}\right)\right]\right]^{1\over3}.
\end{eqnarray}
Using
$\alpha=e^{i\theta} \alpha,$
 we get
\begin{eqnarray}
S_{1}(t)_{cs}&=&\left[\frac{6\pi}{m^{2}_{p}}(|\alpha|^{2}+{1\over2})mt^{2}(1+\frac{1}
{2m^{2}t^{2}})-\frac{3\alpha^{2}}{m}\frac{\pi t^{2}}{m^{2}_{p}}
\right.\nonumber\\
&&\times\left. \left[\frac{\cos2(\theta-mt)}{t^{2}}-\frac{2m}{t} \sin2(\theta-mt)
\right]\right]^{1\over3}.
\end{eqnarray}
Using the approximation ansatzs (32) and the next order perturbative solution
for the squeezed vacuum state becomes,
\begin{eqnarray}
S_{1}(t)_{svs}&=&\left[\frac{6\pi}{m^{2}_{p}}(\sinh^{2}r+{1\over2})mt^{2}
(1+\frac{1}{2m^{2}t^{2}})
+ \frac{6\pi t^2}{m^2_p}\frac{\sinh 2r}{4}\right.\nonumber\\
&&\times\left.\left[\frac{\cos(\varphi-2mt)}{mt^2}
-{2\over t} \sin (\varphi-2mt)\right]\right]^
{1\over 3}.
\end{eqnarray}
Similarly, we get the next order perturbative solution for the squeezed state as
\begin{eqnarray}
S_{1}(t)_{ss}&=&\left[\frac{6\pi}{m^{2}_{p}}(\sinh^{2}r+|\alpha|^2+{1\over2})mt^{2}
(1+\frac{1}{2m^{2}t^{2}})
+ \frac{6\pi t^2}{m^2_p}\frac{\sinh 2r}{4}\right.\nonumber\\
&&\times\left[\frac{\cos(\varphi-2mt)}{mt^2}
-{2\over t} \sin (\varphi-2mt)\right]\nonumber\\
&&-\left.\frac{3\alpha^{2}}{m}\frac{\pi t^{2}}{m^{2}_{p}}
 \left[\frac{\cos2(\theta-mt)}{t^{2}}-\frac{2m}{t} \sin2(\theta-mt)
\right]\right]^{1\over3}.
\end{eqnarray}
From (36),  (37) and (38), it follows that $S_1(t)_{cs}\sim t^
{2\over 3}$, $S_1(t)_{svs}\sim t^{2\over 3}$ and $S_1(t)_{ss}\sim
t^{2\over 3}$

Therefore, in the oscillating phase of the inflaton in coherent,
squeezed vacuum  and squeezed states, the
approximate leading solution to the semiclassical Einstein equation has the
same power-law expansion.
\section{Density fluctuations and validity of semiclassical theory}
The semiclassical theory of gravity valid only if the fluctuations in the
energy momentum tensor are not large. Thus the validity of the semiclassical
theory can be studied with the help of the energy momentum tensor of the inflaton in a suitable
quantum states  under the consideration. We examine the validity of the Einstein equation in squeezed vacuum state which is purely exhibit quatum features. The checking of the validity of the Einstein equation can be done with the help a quantity defined below in terms of the energy momentum tensor as
\begin{equation}
 \Delta=\left|{\left\langle :T^2_{\mu \nu}:\right\rangle -\left\langle :T_{\mu\nu}:\right\rangle ^2 }\right|.
\end{equation}
Where $ \left\langle :T^2_{\mu \nu}:\right\rangle $ is the  expectation value of  the  squared energy momentum tensor of a scalar field in a suitable quatum state and $\left\langle :T_{\mu\nu}:\right\rangle ^2  $is its square of the expectation value.  The meaning of : : is that the expectation values are to be computed with respect to  the normal ordering  of the scalar field. If the $ \Delta$ is more than one then the corresponding semiclassical Einstein equation does not hold in that particular quantum  state.
For the sake of simplicity of the study, we focus on the temporal
component of the energy momentum tensor with single mode of the inflaton and
investigate the validity of the semiclassical theory  in squeezed vacuum
state. Thus we need  consider only the density fluctuations which can be written in squeezed vacuum state as
\begin{equation}
\Delta_{svs} = |<:T^2_{0 0}:>_{svs} - <:T_{0 0}:>^2_{svs}|.
\end{equation}
The first term of eqn(40) can be obtained by squaring  eq(7) and
taking the expectation value in squeezed vacuum state as
\begin{align}
<:T^2_{0 0}:>_{svs} &= {1\over4S^6(t)}<:\hat{\pi}^4:>_{svs} +
{m^2\over4}<:\hat{\pi}^2\hat{\phi}^2:>_{svs} \nonumber\\
&+{m^2\over4}<:\hat{\phi}^2\hat{\pi}^2:>_{svs}+{m^4\over4}S^6(t)<:\hat{\phi}^4:>_{svs}.
\end{align}
Using the eqns (14),(15)  and apply the approximations ansatz then,
 $<:\hat{\pi}^4:>_{svs}$,
$<:\hat{\pi}^2\hat{\phi}^2:>_{svs}$,$<:\hat{\phi}^2\hat{\pi}^2:>_{svs}$
and $<:\hat{\phi}^4:>_{svs}$ can be calculated and are respectively obtained as follows
\begin{eqnarray}
<:\hat{\pi}^4:>_{svs}
&=&S_0^6\biggl[{1\over4m^2}\bigl\{3+6\cosh^2r\sinh^2r+24\sinh^3r\cosh
r\nonumber\\
&&+12(\sinh^2r+\sinh^4r+\cosh r\sinh r)\bigr\}\biggr] \\
<:\hat{\pi}^2\hat{\phi}^2:>_{svs} &= &{1\over4m^2t^2}
\bigl\{3+6\cosh^2r\sinh^2r+24\sinh^3r\cosh
r\nonumber\\
&&+12(\sinh^2r+\sinh^4r+\cosh r \sinh r)\} \\
<:\hat{\phi}^2\hat{\pi}^2:>_{svs} &=&
{1\over4m^2t^2}\biggl\{3+6\cosh^2r\sinh^2r+24\sinh^3r\cosh r\nonumber\\
&&+ 12(\sinh^2r+\sinh^4r+\sinh r\cosh r)\biggr\} \\
<:\hat{\phi}^4:>_{svs} &=&
{1\over4S^6_0t^4m^2}\biggl\{3+6\cosh^2r\sinh^2r+24\cosh r
\sinh^3r\nonumber\\
&&+ 12(\sinh^2r+\sinh^4r+\sinh r\cosh r)\biggr\}.
\end{eqnarray}
 Thus
\begin{align}
<:T^2_{0 0}:>_{svs} & =
\left({1\over16m^2t^4}+{1\over8t^2}+{m^2\over16}\right)
\biggl[3+6\cosh^2r\sinh^2r+24\sinh^3r\cosh r\nonumber\\
&+12\bigl(\sinh^2+\sinh^4r+\cosh r\sinh r\bigr)\biggr]   .
\end{align}
By calculating the expectation value of normal ordered  time-time component of energy momentum
tensor of the inflaton in squeezed vacuum state  and then square the result which leads to
\begin{align}
<:T_{0 0}:>^2_{svs}
&=S^6\biggl[(\sinh^4r+{1\over4}+\sinh^2r)(\dot{\phi}^{*2}\dot{\phi}^2+2m^2
\dot{\phi}^*\dot{\phi}\phi^*\phi+m^4\phi^{*2}\phi^2)\nonumber\\
&+{1\over4}(e^{-2i\varphi}\sinh^2r\cosh^2r)(\dot{\phi}^{*4}+2m^2\dot{\phi}^{*2}\phi^{*2}+m^4\phi^{*4})\nonumber\\
&+{1\over4}e^{2i\varphi}\sinh^2r\cosh^2r(\dot{\phi}^4+m^4\phi^4+2m^2\dot{\phi}^2\phi^2)\nonumber\\
&+\left(\sinh^2r+{1\over2}\right)e^{-i\varphi}\sinh r\cosh
r(\dot{\phi}\dot{\phi}^{*3}+m^2\phi^*\phi\dot{\phi}^{*2}\nonumber\\
&~~~+m^2\dot{\phi}^*\dot{\phi}\phi^{*2}+m^4\phi\phi^{*3})\nonumber\\
&+{1\over2}\sinh^2r\cosh^2r(\dot{\phi}^{*2}\dot{\phi}^2+m^2\dot{\phi}^{*2}\phi^2+m^2
 \phi^{*2}\dot{\phi}^2+m^4\phi^{*2}\phi^2)\nonumber\\
&+ e^{i\varphi}\sinh r\cosh
 r\left(\sinh^2r+{1\over2}\right)(\dot{\phi}^2\dot{\phi}^*\dot{\phi}
 +m^2\phi^2\dot{\phi}^*\dot{\phi}\nonumber\\
& +m^2\dot{\phi}^2\phi^*\phi+m^4\phi^2\phi^*\phi)\biggr].
\end{align}
Using equations (25) and (28) apply the approximation ansatz eqn(32) in (47) we get
\begin{align}
<:T_{0 0}:>^2_{svs}
&=\left({1\over4m^2t^4}+{1\over2t^2}+{m^2\over4}\right)
\biggl({1\over4}+\sinh^4r+\sinh^2r+\sinh^2r\cosh^2r \nonumber\\
&~~~~~+2\sinh^3r\cosh r+\sinh r\cosh r\biggr).
\end{align}
Substituting equations(46) and (48) in eqn (40), then 
\begin{align}
\Delta_{svs} & = \left({1\over16m^2t^4} +
{1\over8t^2}+{m^2\over16}\right)
\biggl[2+2\cosh^2r\sinh^2r+16\sinh^3r\cosh r\nonumber\\
&~~+8(\sinh^2r+\sinh^4r+\cosh r\sinh r)\biggr]     .
\end{align}
In order to examine the validity of the semiclassical equation, we study $\Delta_{svs}$ numerically with the associated squeezing parameter values which are much smaller and larger compared to unity and corresponding results are tabulated in Table I.

In  the  large squeezing limit  compared to much smaller value of the squeezing parameter, the equation(49) becomes
\begin{align}
\Delta_{svs}\approx \left({1\over16m^2t^4} +
{1\over8t^2}+{m^2\over16}\right)
\biggl[{1\over2} + {13\over8}e^{4r}-e^{2r}\biggr].
\end{align}
\begin{table}[]
{\begin{tabular}{cccccccc } 
$r$ & $\Delta_{svs}$ & $r$ & $\Delta_{svs}$ & $r$ & $\Delta_{svs}$ & $r$ & $\Delta_{svs}$\\ 
 0.001  & 0.4970 & 0.010 & 0.4699 & 0.100 & 0.1869 & 1.100 & 32.5227\\
 0.002  & 0.4940 & 0.020 & 0.4397 & 0.200 & 0.1872 & 1.200 & 48.7983\\
 0.003  & 0.4910 & 0.030 & 0.4093 & 0.300 & 0.6828 & 1.300 & 73.0775\\
 0.004  & 0.4880 & 0.040 & 0.3787 & 0.400 & 1.3803 & 1.400 & 109.2970\\
 0.005  & 0.4850 & 0.050 & 0.3477 & 0.500 & 2.3928 & 1.500 & 163.3297\\
 0.006  & 0.4820 & 0.060 & 0.3165 & 0.600 & 3.8845 & 1.600 & 243.9366\\
 0.007  & 0.4790 & 0.070 & 0.2848 & 0.700 & 6.0972 & 1.700 & 364.1878\\
 0.008  & 0.4760 & 0.080 & 0.2527 & 0.800 & 9.3898 & 1.800 & 543.5814\\
 0.009  & 0.4730 & 0.090 & 0.2200 & 0.900 & 14.2961 & 1.900 & 811.2051\\
 0.010  & 0.4699 & 0.100 & 0.1869 & 1.000 & 21.6117 & 2.000 & 1210.4528\\ 
\end{tabular}}

\vspace{.2cm}{\small Table I
Numerical values of $\Delta_{svs}$   for  squeezing parameter $r$  much smaller and larger compared  to unity }
\end{table}
  
\begin{center}
\begin{figure}
{ \includegraphics[scale=0.7]{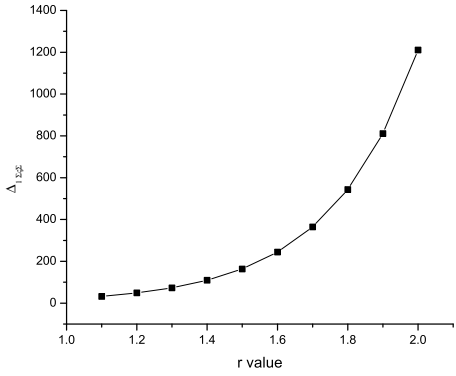}} \hspace{.5cm}
\vspace*{8pt}
\caption{Behaviour of  $\Delta_{svs}$ for  large squeezing limit  compared to unity for the calculated values of  $\Delta_{svs}$ verses  squeezing parameter r\label{f1}}
\end{figure}
\end{center}

 From these analysis, it is clear that the semiclassical Einstein equation in the oscillatory 
phase of the non-classical inflaton in squeezed vacuum state formalism 
is valid only provided the corresponding squeezing parameter  is much smaller compared to unity. 
The  density fluctuations arise  because of the particle creation  due 
to the quantum nature of the squeezed vacuum state  during the oscillatory phase of the inflaton, in the semiclassical gravity.

\begin{center}
\begin{figure}
{\includegraphics[scale=0.7]{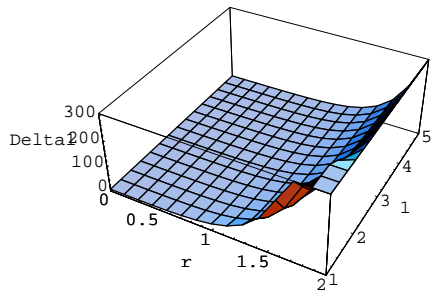}} \hspace{.5cm}
\vspace*{8pt}
\caption{$\Delta_{svs}$   in squeezed vacuum  for  $ r $ and $t$\label{f2}}
\end{figure}
\end{center}

\section{Quantum fluctuations  }
In this section we discuss  quantum fluctuations  for
the coherently oscillating inflaton in semiclassical theory of gravity. The
quantum fluctuations in  coherent  and squeezed states representation
of the inflaton can be studied by using dispersion relations of $\phi$ and
$\pi$, which  are, respectively, given by
\begin{equation}
\left(\Delta \phi\right)^{2}=\langle \hat{\phi}^{2}\rangle -\langle
\hat{\phi}\rangle^{2},
\end{equation}
and
\begin{equation}
\left(\Delta \pi\right)^{2}=\langle \hat{\pi}^{2}\rangle -\langle
\hat{\pi}\rangle^{2}.
\end{equation}
Here, $\langle \hat{\phi}^{2}\rangle$ , $\langle \hat{\pi}^{2}\rangle$,
$\langle \hat{\phi}\rangle$ and  $\langle \hat{\pi}\rangle$ are
respectively the squared expectation values and expectation values of  $\phi$
and $\pi$ in a given quantum
state under consideration.

First, we consider  quantum fluctuations of inflaton in coherent state.
Using (8-11) and (19), it follows that
\begin{equation}
\left(\Delta \phi\right)_{cs}=\sqrt{\phi^{*}\phi},
\end{equation}
and
\begin{equation}
\left(\Delta \pi\right)_{cs}=S^{3}\sqrt{\dot{\phi^{*}}\dot{\phi}}.
\end{equation}
Using eqns( 27) and (28) in the  equations(53) and (54) respectively and then applying the approximation ansatz (32) we get
   the dispersion relations in coherent state  as
 \begin{equation}
\left(\Delta \phi\right)_{cs}=\sqrt{\frac{1}{ S_{0}^{3}t^{2}}{1\over2m}},
\end{equation}
and
 \begin{equation}
\left(\Delta \pi\right)_{cs}=S_{0}^{3}t^{2}\sqrt{\frac{1}{ S_{0}^{3}t^{2}}
{1\over2m}}
\left({1\over t^{2}}+m^{2}\right).
\end{equation}
In the limit $ mt>>1 $,
the above two expressions gives the dispersion relation in coherent states as
 \begin{equation}
\left(\Delta \phi\right)_{cs}\left(\Delta \pi\right)_{cs}={1\over2}\sqrt{
\left(1+{1\over m^{2} t^{2}}\right)}.
\end{equation}
 The dispersion relations in squeezed  vacuum state can be calculated
by the same procedure and is   obtained in the limit $mt > > 1$, as
\begin{eqnarray}
\left(\Delta\phi\right)_{svs}
\left(\Delta\pi\right)_{svs}
&=&{1\over2m}\left[\left[
\left(2\sinh^{2}r+1 \right)
+2\sinh r\cosh r \cos(\varphi-2mt)\right]\right.  \nonumber\\
&\times& \left(2\sinh^{2}r+1 \right)
\left({1 \over t^2}+m^{2}\right)
+\sinh r\cosh r \nonumber\\
&\times&\left. \left[\frac{2}{t^2} \cos(\varphi-2mt)-{4m \over t}
\sin(\varphi-2mt)
-2m^2 \cos(\varphi-2mt)\right]\right]^{1\over2}\,.
\end{eqnarray}
Similarly for the  squeezed state, we get
\begin{eqnarray}
\left(\Delta\phi\right)_{ss}
\left(\Delta\pi\right)_{ss}
&=&{1\over2m}\left[\left[
\left(2\sinh^{2}r+1 \right)
+2\sinh r\cosh r \cos(\varphi-2mt)\right]\right.  \nonumber\\
&\times& \left(2\sinh^{2}r+1 \right)
\left({1 \over t^2}+m^{2}\right)
+\sinh r\cosh r \nonumber\\
&\times&\left. \left[\frac{2}{t^2} \cos(\varphi-2mt)-{4m \over t}
\sin(\varphi-2mt)
-2m^2 \cos(\varphi-2mt)\right]\right]^{1\over2}\,.
\end{eqnarray}
\section{Discussions and Conclusions}
Quantum effects of  matter field can play a significant role in the
early universe. In order to study quantum effects in an early universe scenario, both
background metric and matter field under consideration are to be treated
quantum mechanically. Although, there are have been attempts to formulate the quantum
theory of gravity, still we do not have a consistent quantum theory of
gravity. Therefore appropriate conditions, quantum effects in various
cosmological problems  can be studied in a semiclassical approach.
 In this approach, the background metric is treated as classical and the
matter field as quantum mechanical. In the present work, we studied a
homogeneous massive scalar field minimally coupled to the
Friedmann-Robertson-Walker universe in the context of semiclassical
theory of gravity. The quantum effect of the scalar field is
studied by representing the inflaton in various nonclassical states,
such as coherent  and squeezed states. We particularly  addressed  the oscillatory phase of the
inflaton after inflation in coherent and squeezed states formalisms. 

Approximate leading solutions to the Einstein equation, in 
coherent and squeezed state formalisms  are obtained perturbatively.
It is found that the solution for coherent state depends on the
coherent state parameter
and its phase factor, while the solution for squeezed states, depends on the
associated squeezing parameter and squeezing angle.
It can be concluded that in the oscillatory
phase of the inflaton,
in coherent  and squeezed state representations, the approximate
leading solution to the semiclassical Einstein equation has the same power-law
expansion,  as the classical Einstein equation. However, in
the coherent   and squeezed states
formalisms,  the correction terms do not oscillate in contrast with the classical gravity.

 We studied the quantum fluctuations
due to the coherently oscillating inflaton after the inflation in coherent
state  and squeezed state formalisms, in the frame work of semiclassical theory
of gravity. The quantum fluctuations of the inflaton studied in terms of
dispersion relations of $\phi$ and $\pi$ and the study of the  coherent state
shows that the dispersion relations of the inflaton is inversely proportional to
$t$. Also the study reveals that the uncertainty relation for coherent state
does not depend on the coherent state parameter. While the uncertainty
relation  calculated for squeezed vacuum state shows that the relation
depends on the associated squeezing parameter and the squeezing angle.

 We studied density fluctuations in squeezed vacuum state  formalism.
It is found the analytical result for density fluctuations and hence 
examined the validity of the semiclassical Einstein equation.
 The validity of the semiclassical Einstein equation is studied in terms 
of the energy density of the inflaton field in the oscillating phase of inflaton.
 It is observed that the semiclassical equation hold only when the associated squeezing 
parameter is much less than one otherwise the density fluctuates so widely because of the particle creation  due 
to the quantum nature of the squeezed vacuum state  during the oscillatory phase of the inflaton.

\end{document}